\documentclass[12pt]{article}
 \usepackage{graphicx}
 \usepackage[cp1251]{inputenc} 
\begin{document}
 \noindent {\footnotesize\it Astronomy Letters, 2016, Vol. 42, No. 9, pp. 567--582.}
 \newcommand{\dif}{\textrm{d}}

 \noindent
 \begin{tabular}{llllllllllllllllllllllllllllllllllllllllllllll}
 & & & & & & & & & & & & & & & & & & & & & & & & & & & & & & & & & & & & & \\\hline\hline
 \end{tabular}

  \vskip 0.5cm
  \centerline{\large\bf Rotation Curve and Mass Distribution in the Galaxy from}
  \centerline{\large\bf the Velocities of Objects at Distances up to 200 kpc}
  \bigskip
  \centerline{A.T. Bajkova and V.V. Bobylev }
  \bigskip
  \centerline{\small\it Pulkovo Astronomical Observatory, St. Petersburg,  Russia}
  \bigskip
  \bigskip
{\bf Abstract}—Three three-component (bulge, disk, halo) model
Galactic gravitational potentials differing by the expression for
the dark matter halo are considered. The central (bulge) and disk
components are described by the Miyamoto--Nagai expressions. The
Allen--Santill\'an (I), Wilkinson-Evans (II), and
Navarro--Frenk--White (III) models are used to describe the halo.
A set of present-day observational data in the range of
Galactocentric distances $R$ from 0 to 200~kpc is used to refine
the parameters of these models. For the Allen--Santill\'an model,
a dimensionless coefficient $\gamma$ has been included as a
sought-for parameter for the first time. In the traditional and
modified versions, $\gamma=2.0$ and 6.3, respectively. Both
versions are considered in this paper. The model rotation curves
have been fitted to the observed velocities by taking into account
the constraints on the local matter density
 $\rho_\odot=0.1 M_\odot$~pc$^{-3}$ and the force
 $K_{z=1.1}/2\pi G=77 M_\odot $\rm pc$^{-2}$
 acting perpendicularly to the Galactic plane. The
Galactic mass within a sphere of radius 50~kpc,
 $M_G (R\leq50$~kpc$)\approx(0.41\pm0.12)\times10^{12}M_\odot$,
is shown to satisfy all three models. The differences between the
models become increasingly significant with increasing radius $R.$
In model I, the Galactic mass within a sphere of radius 200 kpc at
$\gamma=2.0$ turns out to be greatest among the models considered,
 $M_G (R\leq200$~kpc$)=(1.45\pm0.30)\times10^{12}M_\odot$,
 $M_G (R\leq200$~kpc$)=(1.29\pm0.14)\times10^{12}M_\odot$
 at $\gamma=6.3$, and the smallest
value has been found in model~II,
 $M_G (R\leq200$~kpc$)=(0.61\pm0.12)\times10^{12}M_\odot.$
 In our view,
model III is the best one among those considered, because it
ensures the smallest residual between the data and the constructed
model rotation curve provided that the constraints on the local
parameters hold with a high accuracy. Here, the Galactic mass is
 $M_G (R\leq200$~kpc$)=(0.75\pm0.19)\times10^{12}M_\odot.$
 A comparative analysis with the models by Irrgang et al. (2013),
including those using the integration of orbits for the two
globular clusters NGC 104 and NGC 1851 as an example, has been
performed. The third model is shown to have subjected to a
significant improvement.


\section*{INTRODUCTION}
When the dynamical properties of stars, globular clusters, or
dwarf satellite galaxies of the Milky Way are studied, the
construction of their Galactic orbits plays an important role.
This requires having a good model of the Galactic gravitational
potential. The Galactic rotation curve, the dependence of the
circular rotation velocity of objects $V_{circ}$ on their distance
from the Galactic rotation axis $R,$ provides an observational
basis for the construction of such models.

At present, there are highly accurate velocities and distances
only for the objects lying either close to the Sun or in the inner
Galaxy $(R<R_\odot).$ The Galactic rotation curve at greater
distances from the rotation axis ($R\gg R_\odot$) is known with
huge errors. The line of-sight velocities of thick-disk and halo
objects (carbon stars, horizontal-branch giants, globular
clusters, dwarf satellite galaxies) are used (Sofue 2009, 2012;
Bhattacharjee et al. 2014) in this region, but they have a large
velocity dispersion and large errors in the distance estimates.
Nevertheless, we know that the Galactic rotation curve does not
remain flat but slowly falls off with increasing distance $R.$
According to the calculations of various authors, the Galactic
mass is either
 $M_G (R\leq385~\hbox {kpc})=0.70\times10^{12}M_\odot$ (Sofue 2012), or
 $M_G (R\leq260~\hbox {kpc})=1.37\times10^{12}M_\odot$ (Eadie et al. 2015), or even
 $M_G (R\leq200~\hbox {kpc})=3.0\times10^{12}M_\odot$ (model III in Irrgang et al.
(2013)), i.e., the estimates differ by several times. The mass of
the dark matter halo at such large radii will be a major
contributor to the Galactic mass. For example,
 $M_H(R\leq385~\hbox {kpc})=0.65\times10^{12}M_\odot$ (Sofue 2012),
where $R=385$~kpc is half the distance between the Milky Way and
the Andromeda Galaxy (M31).

 \begin{figure} {\begin{center}
 \includegraphics[width=120mm]{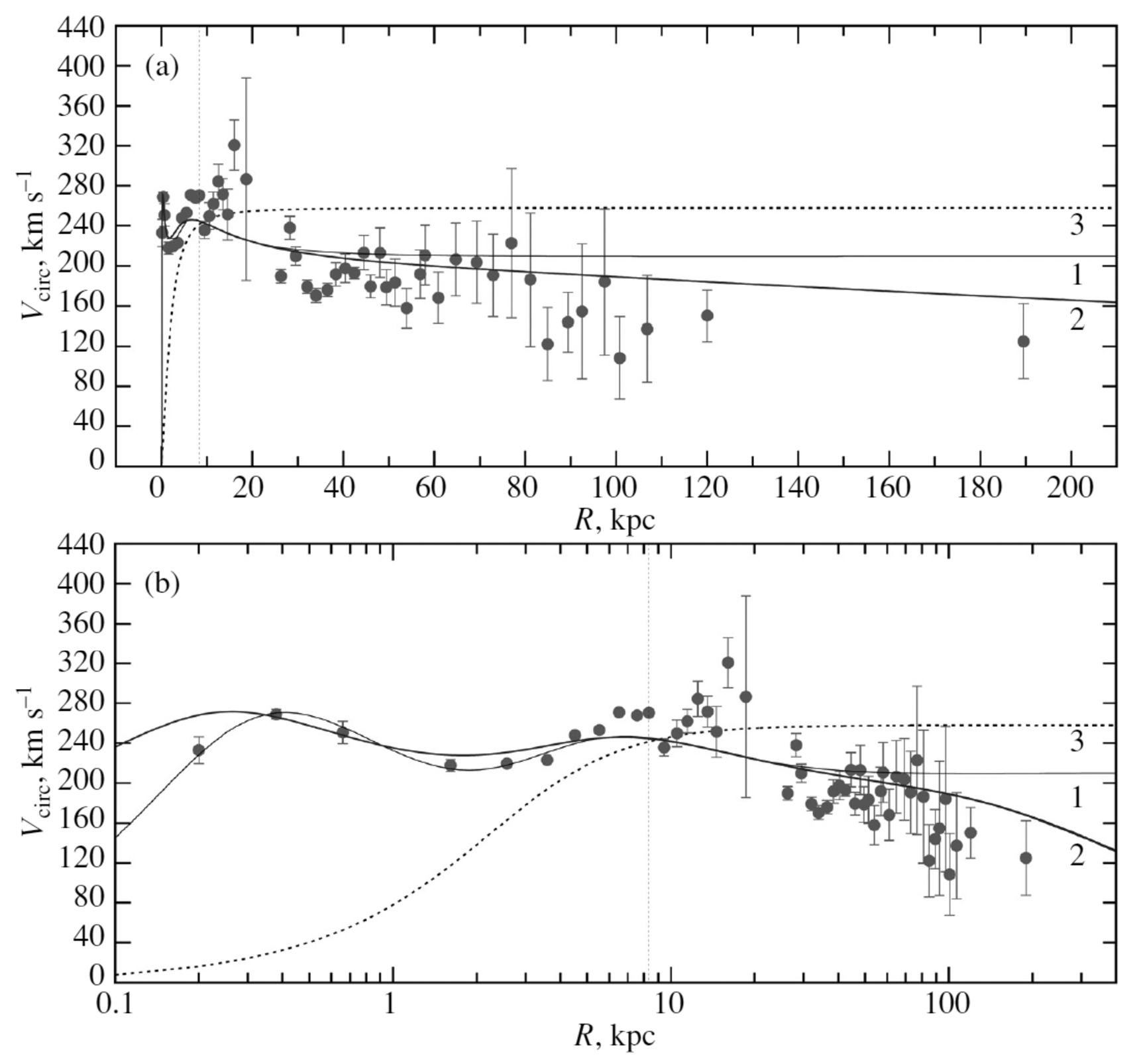}
 \caption{Galactic rotation curve: from Bhattacharjee et al. (2014),
the filled circles with error bars; from Bobylev and Bajkova
(2013), line~1; model~II from Irrgang et al. (2013), line~2; from
Gromov et al. (2015), line~3. The linear and logarithmic distance
scales are on panels (a) and (b), respectively; the vertical line
marks the Sun’s position.}
 \label{f1} \end{center} } \end{figure}

The three-component model potential by Allen and Santill\'an
(1991) is well known. The orbits of various Galactic objects were
determined by various authors using this model. For example, the
orbits of dwarf galaxies (L\'epine et al. 2011), globular clusters
(Lane et al. 2012), planetary nebulae (Wu et al. 2011), and
various stars (Edelmann et al. 2005; Pauli et al. 2006; Pereira et
al. 2012) were determined. A number of modifications of the model
by Allen and Santill\'an (1991), which differ by the halo shape
(Navarro et al. 1997; Wilkinson and Evans 1999) or by the number
of components (Gardner et al. 2011), are known.

Irrgang et al. (2013) and Bobylev and Bajkova (2013) refined the
parameters of the Allen-Santill\'an model using the line-of-sight
velocities of hydrogen clouds at the tangent points and the space
velocities of Galactic masers with measured trigonometric
parallaxes. The sample of masers with measured parallaxes
currently provides the best basis for studying the Galactic thin
disk (Reid et al. 2014). This is possible owing to the high
accuracy of distance determinations, which is, on average, about
10\%.

Figure 1 presents the Galactic rotation curve (in the form of
averaged velocities) from Bhattacharjee et al. (2014), the two
rotation curves found using masers from Bobylev and Bajkova (2013)
and Irrgang et al. (2013), and the rotation curve found from
hydrogen clouds by Gromov et al. (2015) using a quasiisothermal
model potential. Out of the three models in Irrgang et al. (2013),
we took model II where the rotation velocities fall off with
distance faster than in the remaining models proposed by them. All
three curves 1, 2, and 3 were found based on samples of objects
located at distances $R<20$~kpc.

As can be seen from the lower panel in Fig. 1, curves 1 and 2 are
in good agreement with the observational data in the inner Galaxy
$(R<R_\odot).$ All three curves show considerable disagreement
with the data at distances R greater than 30--40 kpc, which is
clearly seen on both panels of Fig.~1. Based on this figure, we
can reach an obvious conclusion about the necessity of improving
the model potentials at great distances.

The goal of this paper is to refine the parameters of the model
Galactic potentials using the Galactic rotation curve in the range
0--200 kpc. We consider three three-component axisymmetric model
potentials (consisting of a bulge, a disk, and a dark matter halo)
that differ by the halo shape, just as was considered previously
by Irrgang et al. (2013). The practical considerations associated
with the numerical integration of orbits, namely the mathematical
simplicity and closeness of the analytical expressions ensuring a
high speed of computations, which is especially important in
implementing the Monte Carlo method to estimate the errors, served
as a justification for the choice of models provided that they
were all capable of reproducing well the observational data.
Whereas Irrgang et al. (2013) used only the masers with measured
trigonometric parallaxes up to distances $R<20$~kpc as
observational data, we add the velocities of objects considerably
farther from the Galactic center to the data on masers. More
specifically, at $20<R<200$~kpc we use the data from Bhattacharjee
et al. (2014), in which almost all of the present-day kinematic
measurements of objects located at great distances are reflected.

 \section*{DATA}
Bhattacharjee et al. (2014) constructed the Galactic rotation
curve in the range of Galactocentric distances 0--200 kpc using
various kinematic data. The line-of-sight velocities of hydrogen
clouds at the tangent points were taken in the inner Galaxy
($R\leq R_\odot$~kpc). These were the data on planetary nebulae,
open star clusters, Cepheids, and carbon stars at distances up to
$R\approx20$~kpc, while the line-of-sight velocities of thick-disk
and halo objects were used at distances up to $R\approx200$~kpc:
1457 blue horizontal branch giants, 2227 K giants, 16 globular
clusters, 28 distant halo giants, and 21 dwarf galaxies. It is
important to note that Bhattacharjee et al. (2014) constructed the
Galactic rotation curve with $R_\odot=8.3$~kpc and
$V_\odot=244$~km s$^{-1}$. Estimates close to these values were
obtained by Irrgang et al. (2013) when analyzing the masers with
measured trigonometric parallaxes for three model potentials. Reid
et al. (2014) estimated $R_\odot=8.34\pm0.16$~kpc and
$V_\odot=240\pm8$~km s$^{-1}$ by analyzing the kinematics of
Galactic masers, while Bobylev and Bajkova found
$R_\odot=8.3\pm0.2$~kpc and $V_\odot=241\pm7$~km s$^{-1}$ from
masers.

It is clearly seen from Fig. 1 that a number of points on the
rotation curve from Bhattacharjee et al. (2014) near
$R\approx20$~kpc have a large dispersion and show an abrupt jump
into the region of high velocities. We decided not to use these
data, especially since Bhattacharjee et al. (2014) used the
kinematic distance estimates (for example, for HII regions from
Hou et al. (2009) and Urquhart et al. (2012)) for a number of
objects from this range of distances.

As a result, we rely on the line-of-sight velocities of hydrogen
clouds at the tangent points and the data on 130 masers with
measured trigonometric parallaxes at distances $R$ less than
25~kpc, and the rotation curve from Bhattacharjee et al. (2014)
serves as the data at greater distances.

The following can be said in more detail about the sample of
masers with measured trigonometric parallaxes. The VLBI
measurements of 103 masers are described in the review of Reid et
al. (2014). A number of publications by these authors devoted to
the analysis of masers located in individual Galactic spiral arms
appeared subsequently, with improved values of the measured
parameters having been given for some of the masers. The papers by
Wu et al. (2014), Choi et al. (2014), Sato et al. (2014),
Hachisuka et al. (2015), and Sanna et al. (2014) are devoted to
the analysis of masers in the Carina--Sagittarius arm, the Perseus
arm, the Scutum arm, the Outer Arm, and the central region of the
Galaxy, respectively; finally, the paper by Xue et al. (2013)
devoted to the masers in the Local Arm was published slightly
earlier.

After the addition of the most recent astrometric measurements of
masers (Motogi et al. 2015; Burns et al. 2015), we obtained a
sample containing the data on 130 sources. We did not include the
masers located at distances $R<4$~kpc in our sample, because the
maser velocities here have large dispersions. The data on hydrogen
clouds are better suited for this region. We use the line-of-sight
velocities of HI clouds at the tangent points from the central
region of the Galaxy (Burton and Gordon 1978). Note, finally, that
the circular rotation velocities $(V_{circ})$ of masers were
derived from their total space velocities, which increases
considerably the reliability of these data. In contrast, when
constructing the velocities of many other objects at distances
$R>25$~kpc, we used only their line-of-sight velocities.

 \section*{MODEL POTENTIALS}
 \subsection*{Introductory Concepts}
In all of the models here, the axisymmetric Galactic potential is
represented as a sum of three components—a central spherical bulge
$\Phi_b(r(R,z))$, a disk $\Phi_d(r(R,z))$, and a massive spherical
dark matter halo $\Phi_h(r(R,z))$:
 \begin{equation}
  \Phi(R,z)=\Phi_b(r(R,z))+\Phi_d(r(R,z))+\Phi_h(r(R,z)).
 \label{pot}
 \end{equation}
We use a cylindrical coordinate system ($R,\psi,z$) with the
coordinate origin at the Galactic center. In a rectangular
coordinate system $(x,y,z)$ with the coordinate origin at the
Galactic center, the distance to a star (spherical radius) will be
$r^2=x^2+y^2+z^2=R^2+z^2$.

In accordance with the convention adopted in Allen and Santill\'an
(1991), we express the gravitational potential in units of 100
km$^2$~s$^{-2}$, the distances in kpc, and the masses in units of
the Galactic mass $M_{gal}=2.325\times 10^7 M_\odot$,
corresponding to the gravitational constant $G=1.$

The expression for the mass density follows from the Poisson
equation
\begin{equation}
4\pi G\rho(R,z)=\nabla^2\Phi(R,z) \label{pois1}
\end{equation}
and is
\begin{equation}
 \rho(R,z)=\frac{1}{4\pi G}\Bigg{(}
 \frac{d^2\Phi(R,z)}{dR^2}+\frac{1}{R}\frac{d\Phi(R,z)}{dR}+
 \frac{d^2\Phi(R,z)}{dz^2}\Bigg{)}.
\label{pois2}
\end{equation}
The force acting in the $z$ direction perpendicularly to the
Galactic plane is expressed as
\begin{equation}
K_z(z,R)=-\frac{d\Phi(z,R)}{dz}. \label{Kz}
\end{equation}
We will need Eqs. (3) and (4) below to solve the problem of
fitting the parameters of the model gravitational potentials with
constraints imposed on the local dynamical mass density
$\rho_\odot$ and the force $K_z(z,R_\odot)$~kpc, which are known
from observations. In addition, we will need the expressions to
calculate:

(1) the circular velocities
\begin{equation}
V_{circ}(R)=\sqrt{R\frac{d\Phi(R,0)}{dR}}, \label{V}
\end{equation}

(2) the Galactic mass contained in a sphere of radius $r$
\begin{equation}
m(<r)=r^2\frac{d\Phi(r)}{dr}, \label{m}
\end{equation}

(3) the parabolic velocity or the escape velocity of a star from
the attractive Galactic field
\begin{equation}
V_{esc}(R,z)=\sqrt{-2\Phi(R,z)}, \label{Vesc}
\end{equation}

(4) the Oort parameters
\begin{equation}
A=\frac{1}{2}R_\odot\Omega_\odot^{'}, \label{A}
\end{equation}
\begin{equation}
B=\Omega_\odot+A, \label{B}
\end{equation}
where $\Omega=V/R$ is the angular velocity of Galactic rotation
$(\Omega_\odot=V_\odot/R_\odot)$, $\Omega^{'}$ is the first
derivative of the angular velocity with respect to $R,$ and
$R_\odot$ is the Galactocentric distance of the Sun.

(5) the surface density of gravitating matter within $z_{out}$ of
the Galactic $z=0$ plane
\begin{equation}
 \Sigma_{out}(z_{out})=2\int_0^{z_{out}} \rho(R,z)dz=
 \frac{K_{z}}{2\pi G}+\frac{2z_{out}(B^2-A^2)}{2\pi G},
\label{Sigma}
\end{equation}

 \subsection*{Bulge and Disk}
In all of the models being considered here, the bulge,
$\Phi_b(r(R,z))$, and disk, $\Phi_d(r(R,z))$, potentials are
represented in the form proposed by Miyamoto and Nagai (1975):
 \begin{equation}
  \Phi_b(r)=-\frac{M_b}{(r^2+b_b^2)^{1/2}},
  \label{bulge}
 \end{equation}
 \begin{equation}
 \Phi_d(R,z)=-\frac{M_d}{\{R^2+[a_d+(z^2+b_d^2)^{1/2}]^2\}^{1/2}},
 \label{disk}
\end{equation}
where $M_b$ and $M_d$ are the masses of the components, $b_b,
a_d,$ and $b_d$ are the scale lengths of the components in kpc.
The corresponding expressions for the mass densities $\rho_b(R,z)$
and $\rho_d(R,z)$ are
\begin{equation}
\rho_b(r)=\frac{3b_b^2 M_b}{4\pi(r^2+b_b^2)^{5/2}}, \label{ro-b}
\end{equation}
\begin{equation}
 \rho_d(R,z)=\frac{b_d^2 M_d}{4\pi(z^2+b_d^2)^{3/2}}
 \frac{a_d R^2+(a_d+3\sqrt{z^2+b_d^2})(a_d+\sqrt{z^2+b_d^2})^2}
 {(R^2+(a_d+\sqrt{z^2+b_d^2})^2)^{5/2}}.
\label{ro-d}
\end{equation}
Expressions (11) and (13) are called a Plummer (1911) sphere,
while relations (12) and (14) are called a generalized Kuzmin
(1956) disk. Integrating the mass densities over the entire volume
of the Galaxy gives the expected bulge and disk masses: $m_b=M_b,
m_d=M_d$. The contributions of the bulge and the disk to the
circular velocity are, respectively,
\begin{equation}
V_{circ(b)}^2(R)=\frac{M_b R^2}{(R^2+b_b^2)^{3/2}}, \label{Vc-b}
\end{equation}
\begin{equation}
V_{circ(d)}^2(R)=\frac{M_d R^2)}{(R^2+(a_d+b_d)^2)^{3/2}}.
\label{Vc-d}
\end{equation}
The corresponding expressions for $K_z^b(z,R)$ and $K_z^d(z,R)$
are
\begin{equation}
K_{z}^{b}(z,R)=\frac{z M_b}{(R^2+z^2+b_b^2)^{3/2}}, \label{Kz-b}
\end{equation}
\begin{equation}
 K_{z}^{d}(z,R)=\frac{z M_d (a_d+\sqrt{z^2+b_d^2})}
 {\sqrt{z^2+b_d^2}(R^2+(a_d+\sqrt{z^2+b_d^2})^2)^{3/2}}.
\label{Kz-d}
\end{equation}

 \subsection*{Halo}
{\bf Model I.} The expression for the halo potential was derived
by Irrgang et al. (2013) based on the expression for the halo mass
from Allen and Martos (1986):
\begin{equation}
 \renewcommand{\arraystretch}{3.2}
  m_h(<r) = \left\{
  \begin{array}{ll}\displaystyle
  \frac{M_h(r/a_h)^{\gamma}}{1+(r/a_h)^{\gamma-1}},
  & \textrm{if }  r\leq\Lambda \\\displaystyle
  \frac{M_h(\Lambda/a_h)^{\gamma}}{1+(\Lambda/a_h)^{\gamma-1}}=\textrm{const},
  & \textrm{if } r>\Lambda  \end{array} \right\},
 \label{m-h-I}
 \end{equation}
It slightly differs from that given in Allen and Santill\'an
(1991) and is
\begin{equation}
 \renewcommand{\arraystretch}{3.2}
  \Phi_h(r) = \left\{
  \begin{array}{ll}\displaystyle
  \frac{M_h}{a_h}\biggl( \frac{1}{(\gamma-1)}\ln \biggl(\frac{1+(r/a_h)^{\gamma-1}}{1+(\Lambda/a_h)^{\gamma-1}}\biggr)-
  \frac{(\Lambda/a_h)^{\gamma-1}}{1+(\Lambda/a_h)^{\gamma-1}}\biggr),
  &\textrm{if }   r\leq \Lambda \\\displaystyle
  -\frac{M_h}{r} \frac{(\Lambda/a_h)^{\gamma}}{1+(\Lambda/a_h)^{\gamma-1}}, &\textrm{if }  r>\Lambda,
  \end{array} \right.
 \label{halo-I}
 \end{equation}
where $M_h$ is the mass, $a_h$ is the scale length, the
Galactocentric distance is $\Lambda=200$ kpc, and the
dimensionless coefficient $\gamma=2.0$. The mass density is
represented as
\begin{equation}
 \renewcommand{\arraystretch}{2.2}
  \rho_h(r) = \left\{
  \begin{array}{ll}\displaystyle
  \frac{M_h}{4\pi a_h}
  \frac{(r/a_h)^{\gamma-1} ((r/a_h)^{\gamma-1}+\gamma)}
  {r^2\biggl(1+(r/a_h)^{\gamma-1}\biggr)^2}, & \textrm{if } r\leq \Lambda.\\
     0, & \textrm{if } r>\Lambda.
  \end{array} \right.
 \label{ro-h-I}
 \end{equation}
The contribution of the halo to the circular velocity is
\begin{equation}
 \renewcommand{\arraystretch}{3.2}
  V_{circ(h)}^2(R) = \left\{
  \begin{array}{ll}\displaystyle
  \frac{M_h R^{\gamma-1}}{ a_h^{\gamma}\biggl(1+(R/a_h)^{\gamma-1}\biggr)},
  &\textrm{if }   r\leq \Lambda \\\displaystyle
  \frac{M_h}{R} \frac{(\Lambda/a_h)^{\gamma}}{1+(\Lambda/a_h)^{\gamma-1}},
  &\textrm{if }  r>\Lambda.
  \end{array} \right.
 \label{Vc-h-I}
 \end{equation}
The expression for $K_z^h(z,R)$ at $r\leq\Lambda$ is
\begin{equation}
K_z^h(z,R)=\frac{z
M_h(\sqrt{R^2+z^2}/a_h)^{\gamma-1}}{a_h^2\sqrt{R^2+z^2}(1+(\sqrt{R^2+z^2}/a_h)^{\gamma-1})}.
\label{Kz-h-I}
\end{equation}

{\bf Model II.} The halo component is represented in the form
proposed by Wilkinson and Evans (1999) as
 \begin{equation}
  \Phi_h(r)=-\frac{M_h}{a_h} \ln {\Biggl(\frac{a_h+\sqrt{r^2+a^2_h}} {r}\Biggr)}.
 \label{halo-II}
 \end{equation}
The mass density is calculated from the formula
\begin{equation}
  \rho_h(r)= \frac{M_h}{4\pi} \frac{a^2_h} {r^2(r^2+a^2_h)^{3/2}}.
 \label{ro-h-II}
 \end{equation}
The contribution of the halo to the circular velocity is
 \begin{equation}
  V^2_{circ(h)}(R)= \frac{M_h}{\sqrt{R^2+a^2_h}}.
 \label{Vc-h-II}
 \end{equation}
The expression for $K_z^h(z,R)$ is
\begin{equation}
K_z^h(z,R)=\frac{z M_h}{(R^2+z^2)\sqrt{R^2+z^2+a_h^2}}.
\label{Kz-h-II}
\end{equation}

{\bf Model III.} The halo component is represented in the form
proposed by Navarro et al. (1997) as
 \begin{equation}
  \Phi_h(r)=-\frac{M_h}{r} \ln {\Biggl(1+\frac{r}{a_h}\Biggr)}.
 \label{halo-III}
 \end{equation}
This model is often called the NFW (Navarro, Frenk, White) model.
The corresponding mass density is
\begin{equation}
  \rho_h(r)= \frac{M_h}{4\pi} \frac{1} {r(r+a_h)^2}.
 \label{ro-h-III}
 \end{equation}
The contribution to the circular velocity is
 \begin{equation}
  V^2_{circ(h)}(R)= M_h \biggl[\frac{\ln(1+R/a_h)}{R}-\frac{1}{R+a_h}\biggr].
 \label{Vc-h-III}
 \end{equation}
The expression for $K_z^h(z,R)$ is
\begin{equation}
 K_z^h(z,R)=\biggl|\frac{z M_h}{(R^2+z^2)}
 \biggl(\frac{1}{a_h+\sqrt{R^2+z^2}}-\frac{\ln(1+\sqrt{R^2+z^2}/a_h)}
 {\sqrt{R^2+z^2}}\biggr)\biggr|.
\label{Kz-h-III}
\end{equation}

 {\begin{table}[t]                            
 \caption[] {\small\baselineskip=1.0ex
 The parameters of models I--III found by fitting to the data}
 \label{t:1}
 \small \begin{center}\begin{tabular}{|c|r|r|r|r|r|r|}\hline
        Parameters &  Model~I,$\gamma=2.0$& Model~I, $\gamma=6.3$  & Model~II &  Model~III \\\hline
  $M_b$ ($M_{g}$) &       386 $\pm10$     &496$\pm19$   & 142$\pm12$     & 443$\pm27$       \\
  $M_d$ ($M_{g}$) &      3092 $\pm62$     &3200$\pm115$ & 2732$\pm16$    & 2798$\pm84$      \\
  $M_h$ ($M_{g}$) &       452 $\pm83$     &1396$\pm122$ & 24572$\pm5459$ & 12474$\pm3289$   \\
      $b_b$ (kpc) &  0.2487   $\pm0.0060$ &0.2809$\pm0.0077$ & 0.2497$\pm0.0093$ &0.2672$\pm0.0090$ \\
      $a_d$ (kpc) &      3.67 $\pm0.16$   &3.12$\pm0.20$    & 5.16$\pm0.32$     & 4.40 $\pm0.73$   \\
      $b_d$ (kpc) &    0.3049 $\pm0.0028$ &0.2929$\pm0.0046$ & 0.3105$\pm0.0032$ & 0.3084$\pm0.0050$\\
      $a_h$ (kpc) &      1.52 $\pm0.18$   &5.39$\pm0.44$      &   64.3 $\pm15$    & 7.7$\pm2.1$  \\\hline
   Entropy of res. noise $E$   &  $-31.40$ & $-26.51$ & $-27.78$ & $-24.51$ \\\hline
      $\delta $ (km s$^{-1}$)        &     15.7  &   14.1   &   13.8   &   13.1   \\\hline
      $\delta_{irg}$ (km s$^{-1}$)   &     19.4  &     ---  &   16.4   &   38.4   \\
 \hline
 \end{tabular}\end{center}
 \leftline {\small The Galactic mass unit is $M_{g}=2.325\times10^7 M_\odot$.}
 \end{table}}
 {\begin{table}[t]                            
 \caption[] {\small\baselineskip=1.0ex
 The values of the quantities calculated from the parameters of models I--III found}
 \label{t:2}
 \small \begin{center}\begin{tabular}{|c|r|r|r|r|r|r|}\hline
        Parameters &  Model~I, $\gamma=2.0$& Model~I, $\gamma=6.3$  & Model~II &  Model~III \\\hline
  $(\rho_\odot)_d$ ($M_\odot$ pc$^{-3}$)&0.092$\pm0.010$&0.090$\pm0.010$& 0.090$\pm0.010$&0.089$\pm0.011$  \\
  $(\rho_\odot)_h$ ($M_\odot$ pc$^{-3}$)&0.008$\pm0.001$&0.009$\pm0.001$& 0.010$\pm0.001$&0.010$\pm0.001$  \\
  ~~~ $\rho_\odot$ ($M_\odot$ pc$^{-3}$)&0.100$\pm0.010$&0.100$\pm0.010$& 0.100$\pm0.010$&0.100$\pm0.010$  \\
  ~~~ $K_{z=1.1}/2\pi G$ ($M_\odot$ pc$^{-2}$) &77.2$\pm6.9$&77.0$\pm7.1$   &77.01$\pm10.2$  &77.1$\pm12.5$  \\
  ~~~ $\sum_{1.1}$ ($M_\odot$ pc$^{-2}$)&71.4$\pm7.3$&71.9 $\pm8.1$   &75.78$\pm10.1$  &76.8$\pm12.3$  \\
  ~~~ $\sum_{out}$ ($M_\odot$ pc$^{-2}$)&44.73$\pm8.25$&53.8 $\pm10.3$   & 66.7$\pm10.0$  &69.9$\pm17.6$ \\
 $V_{esc, ~R=R_\odot}$ (km s$^{-1}$) & 561.4 $\pm46.5$ &506.8 $\pm21.0$&518.0$\pm56.2$   &537.8$\pm70.1$ \\
 $V_{esc, ~R=200\,\hbox {\footnotesize kpc}}$(km s$^{-1}$)&250.0$\pm25.6$&115.5$\pm5.0$&164.4$\pm16.0$&210.6$\pm26.2$\\
 $V_\odot$ (km s$^{-1}$) &239.0 $\pm12.0$&245.0 $\pm7.6$ &242.5$\pm28.0$&243.9$\pm34.5$ \\
 $A$ (km s$^{-1}$ kpc$^{-1}$)   &16.01$\pm0.80$&15.89$\pm 0.72$&15.11$\pm1.84$&15.04$\pm2.37$ \\
 $B$ (km s$^{-1}$ kpc$^{-1}$)   &-12.79 $\pm1.06$&-13.64 $\pm0.89$&-14.10$\pm1.77$&-14.35$\pm2.12$ \\
 $M_{G_{(R\leq~50\, kpc)}} $($10^{12}M_\odot$)&0.415$\pm 0.074$&0.386 $\pm0.036$&0.416$\pm0.094$&0.406$\pm0.115$ \\
 $M_{G_{(R\leq100\, kpc)}} $($10^{12}M_\odot$)&0.760 $\pm0.149$&0.686 $\pm0.072$&0.546$\pm0.108$&0.570$\pm0.153$ \\
 $M_{G_{(R\leq150\, kpc)}} $($10^{12}M_\odot$)&1.105 $\pm0.224$&0.987 $\pm0.108$&0.591$\pm0.114$&0.674$\pm0.177$\\
 $M_{G_{(R\leq200\, kpc)}} $($10^{12}M_\odot$)&1.450 $\pm0.300$&1.288 $\pm0.144$& 0.609$\pm0.117$&0.750$\pm0.194$\\
 \hline
 \end{tabular}\end{center}
 \end{table}}

\subsection*{Parameter Fitting}
As follows from Bhattacharjee et al. (2014), the velocities of all
objects on the Galactic rotation curve were calculated with
$R_\odot=8.3$~kpc and $V_\odot=244$~km s$^{-1}$. The parameters of
the model potentials I--III are found by least-squares fitting to
the measured circular rotation velocities $(V_{circ})$ of Galactic
objects. We applied the unit weights at which the smallest
residual between the data and the rotation curve was achieved.

The local dynamical mass density $\rho_\odot$, which is the sum of
the bulge, disk, and dark matter densities in a small solar
neighborhood, together with the surface density $\Sigma_{1.1}$ are
the most important additional constraints in the problem of
fitting the parameters of the model potentials to the measured
circular velocities (Irrgang et al. 2013):
\begin{equation}
\rho_\odot=\rho_b(R_\odot)+\rho_d(R_\odot)+\rho_h(R_\odot),
\label{ro}
\end{equation}
\begin{equation}
\Sigma_{1.1}=
  \int\limits^{1.1\,\hbox {\footnotesize\it kpc}}_{-1.1\,\hbox {\footnotesize\it kpc}}
(\rho_b(R_\odot,z)+\rho_d(R_\odot,z)+\rho_h(R_\odot,z))dz.
\label{Sig}
\end{equation}
The surface density is closely related to the force $K_z(z,R)$ in
accordance with Eq. (10). Since the two most important parameters
$\rho_\odot$ and $K_z/2\pi G$ are known from observations with a
sufficiently high accuracy, introducing additional constraints on
these two parameters allows the parameters of the gravitational
potential to be refined significantly.

For example, according to the analysis of the distribution of
stars from the Hipparcos Catalogue (1997) performed by Holmberg
and Flynn (2000), $\rho_\odot=0.102\pm0.010~M_\odot$~pc$^{-3}$c.
Values of $\rho_\odot$ fairly close to this value were obtained
subsequently from various data. For example,
 $\rho_\odot=0.120^{+0.016}_{-0.019}~M_\odot$~pc$^{-3}$ (Garbari et al. 2012),
 $\rho_\odot=0.091\pm0.0056~M_\odot$~pc$^{-3}$ (Bienaym\'e et al. 2014),
or
 $\rho_\odot=0.097\pm0.013 M_\odot$~pc$^{-3}$ (McKee et al. 2015).

The estimates of the local surface density $\Sigma$ are more
difficult to compare, because this quantity depends on the adopted
scale height and the limits of integration, which were taken to be
different by different authors. For example, Korchagin et al.
(2003) estimated $\Sigma_{0.35}=42\pm6~M_\odot$ pc$^{-2}$ (here,
$|z|\leq0.35$~kpc) based on K-type giants from the Hipparcos
Catalogue. Therefore, for a proper comparison of the results, we
should extrapolate to the same range of heights, for example, to
$\pm1.1$~kpc as in Eq. (10).

By analyzing the distribution of K giants in the solar
neighborhood, Holmberg and Flynn (2004) estimated
 $\Sigma_{1.1}=74\pm6~M_\odot$~pc$^{-2}$. Since Irrgang et al.
(2013) relied on this estimate, the three model Galactic
potentials constructed by them from hydrogen and masers have
 $\Sigma_{1.1}=74-75~M_\odot$~pc$^{-2}$. Present-day estimations
give slightly lower values of this parameter. For example, based
on a sample of 9000 K stars with measured spectra from the
SDSS/SEGUE catalogues, Zhang et al. (2013) estimated
 $\Sigma_{1.0}=67\pm6~M_\odot$~pc$^{-2}$; Bovy and Rix (2013) found
 $\Sigma_{1.1}=68\pm4~M_\odot$~pc$^{-2}$ based on
16300 G dwarfs from the SEGUE catalogue. A useful and
comprehensive review of present-day local density determinations
can be found in Reid (2014). In particular, a new summary estimate
of the local surface density of baryons (HI, H$_2$, gas, stars,
stellar remnants/brown dwarfs),
 $\Sigma_{b}=54.2\pm4.9~M_\odot$~pc$^{-2}$, was obtained in this review.

We took the values from Irrgang et al. (2013) as the target
parameters in fitting $\rho_\odot$ and $K_{z=1.1}/2\pi G$. As a
result, we used two additional constraints:

 (i) the local matter density in the Galaxy $\rho_\odot$ must be close to
 $\widetilde{\rho}_\odot=0.10~M_\odot$~pc$^{-3}$ known from observations
 (Holmberg and Flynn 2000);

(ii) the local force acting perpendicularly to the Galactic plane
must be close to
 $\widetilde{K}_{z=1.1}/2\pi G=77~M_\odot$~pc$^{-2}$ at $R=R_\odot$,
 which corresponds to $\Sigma_{1.1}=74~M_\odot$~pc$^{-2}$ found by Holmberg and Flynn (2004)
 from the observations of K giants in the solar neighborhood.

Thus, the parameter fitting problem was reduced to minimizing the
following quadratic functional $F$:
\begin{equation}
\min F=\sum_{i=1}^N (V_{circ}(R_i)-\widetilde{V}_{circ}(R_i))^2
+\alpha_1(\rho_\odot-\widetilde{\rho}_\odot)^2+\alpha_2(K_{z=1.1}/2\pi
G-\widetilde{K}_{z=1.1}/2\pi G)^2, \label{F}
\end{equation}
where $N$ is the number of data points; the tilde denotes the data
from circular velocity measurements; $R_i$ are the Galactocentric
distances of the objects; $\alpha_1$ and $\alpha_2$ are the weight
factors at the additional constraints that were chosen so as to
minimize the residual between the data and the model rotation
curve provided that the additional constraints hold with an
accuracy of at least 5\%. Based on the constructed models, we
calculated the local surface density of the entire matter
$\rho_\odot$ and $K_{z=1.1}/2\pi G$ related to $\Sigma_{1.1}$ and
$\Sigma_{out}.$ The accuracies of all the parameters given in
Tables 1 and 2 were determined through Monte Carlo simulations
using each time 100 independent realizations of random measurement
errors obeying a normal distribution with zero mean and a known
rms deviation.

\begin{figure}[t]
{\begin{center}
   \includegraphics[width=0.99\textwidth]{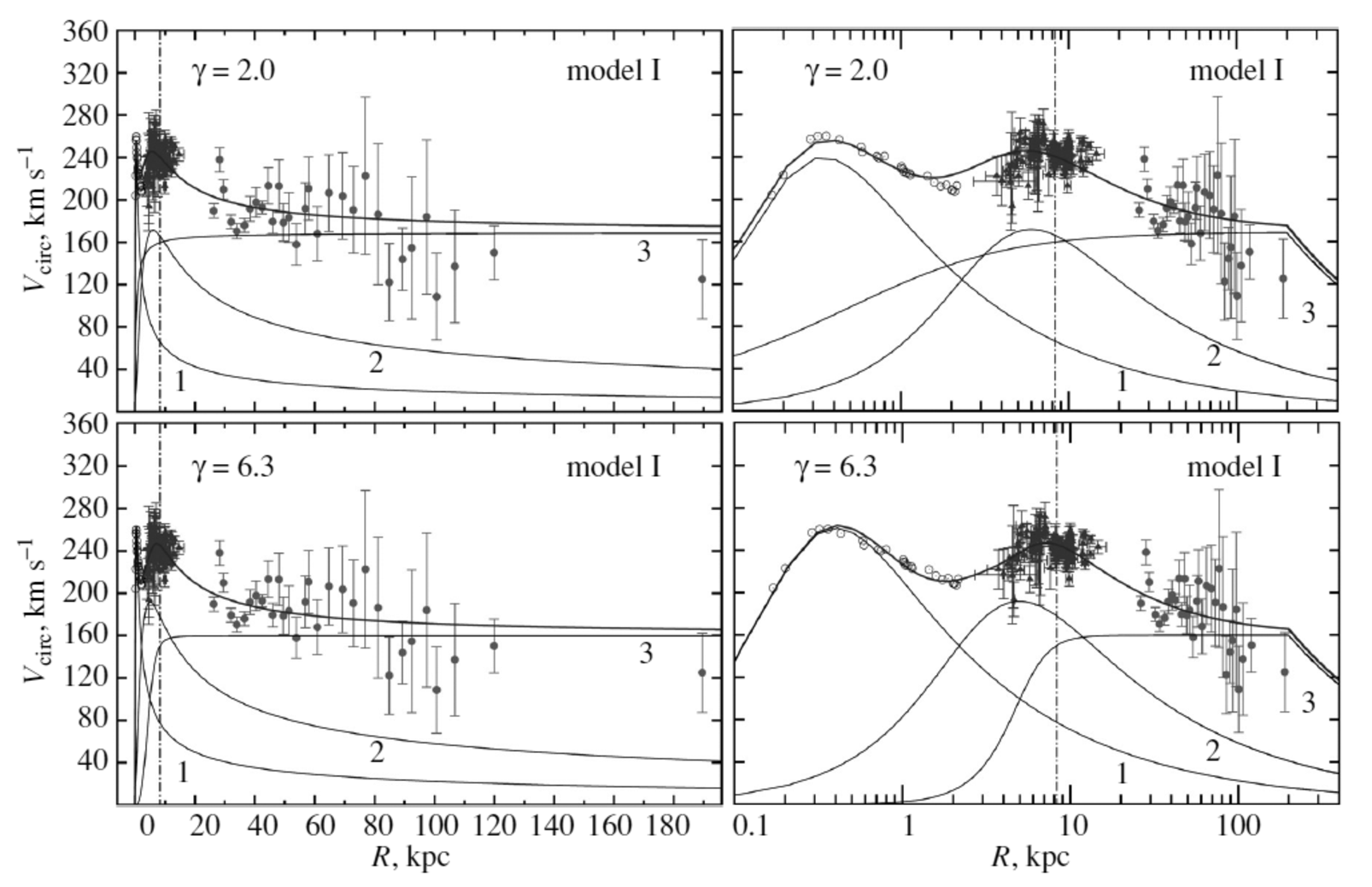}
 \caption{
Galactic rotation curve for model I ($\gamma=2.0$ and 6.3) in
linear (left) and logarithmic (right) distance scales; the
vertical line marks the Sun’s position, numbers 1, 2, and 3 denote
the bulge, disk, and halo contributions, respectively; the open
circles, filled triangles, and filled circles indicate the HI
velocities, the velocities of masers with measured trigonometric
parallaxes, and the velocities from Bhattacharjee et al. (2014),
respectively..
  } \label{f2}
\end{center}}
\end{figure}
\begin{figure}[t]
{\begin{center}
   \includegraphics[width=0.99\textwidth]{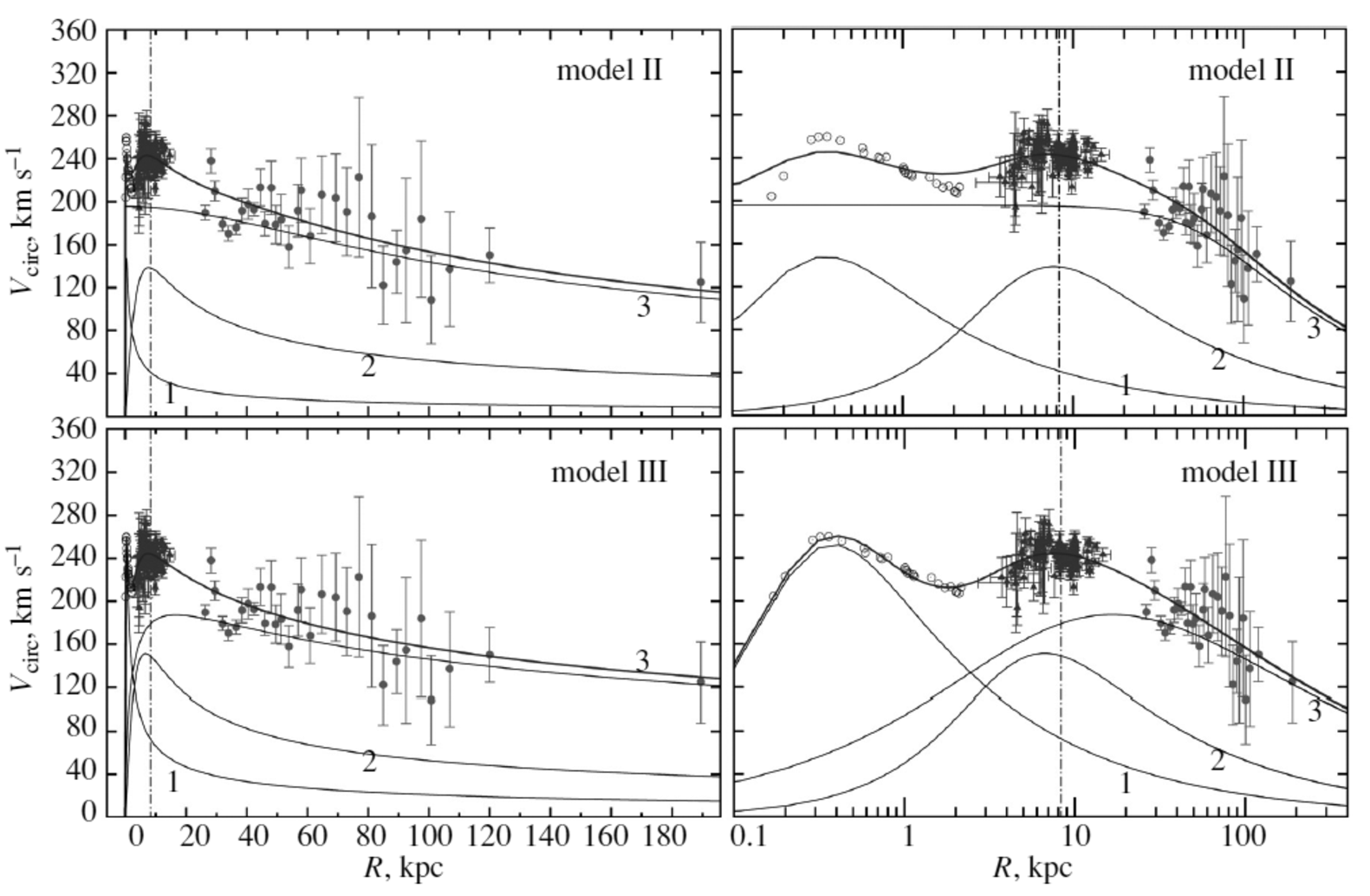}
 \caption{
Galactic rotation curve for models II and III in linear (left) and
logarithmic (right) distance scales; the vertical line marks the
Sun’s position, numbers 1, 2, and 3 denote the bulge, disk, and
halo contributions, respectively; the open circles, filled
triangles, and filled circles indicate the HI velocities, the
velocities of masers with measured trigonometric parallaxes, and
the velocities from Bhattacharjee et al. (2014), respectively.
  } \label{f3}
\end{center}}
\end{figure}
\begin{figure}[t]
{\begin{center}
   \includegraphics[width=0.6\textwidth]{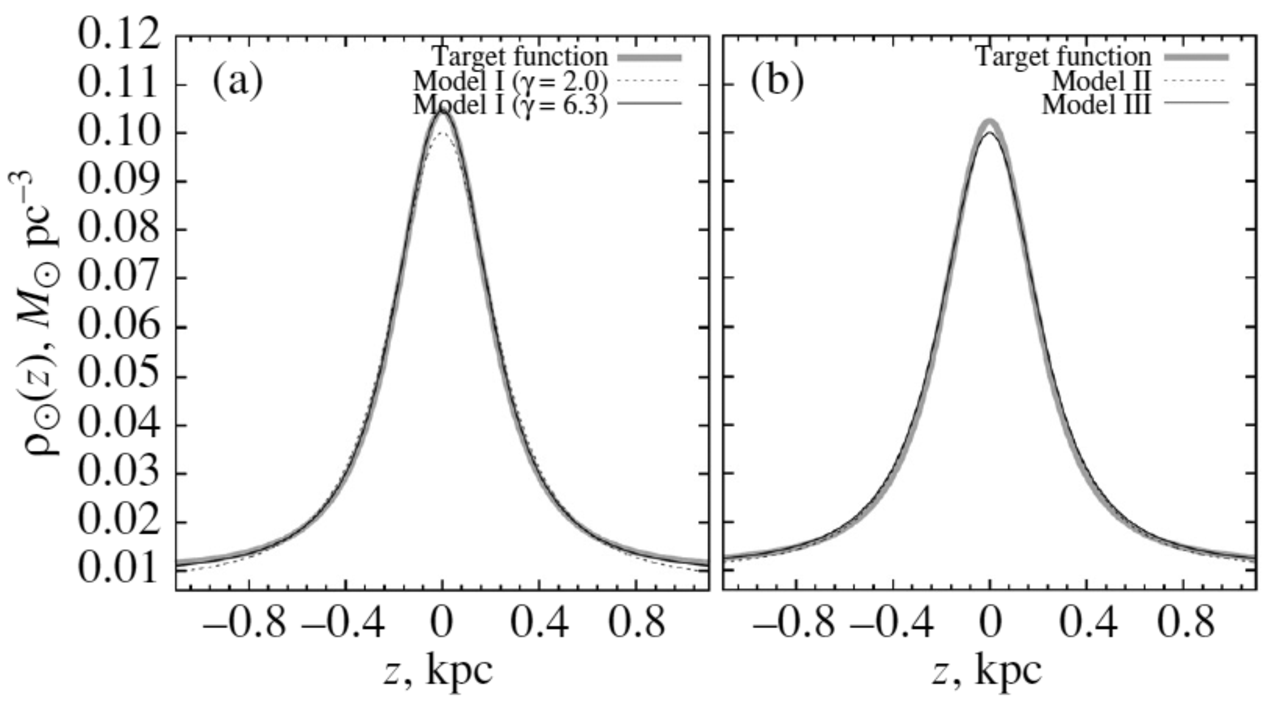}
 \caption{
Results of fitting $\rho_\odot(z)$ for models I, II, and III. The
functions from Irrgang et al. (2013) are considered as the target
function $\rho_\odot(z)$.
  } \label{f4}
\end{center}}
\end{figure}

 \section*{RESULTS}
Table 1 provides the values of the seven sought for parameters
($M_b, M_d, M_h, b_b, a_d, a_b, a_h$) found by solving the fitting
problem for the three model Galactic potentials under
consideration. The next-to-last row in the table gives the
residuals (in km s$^{-1}$) between the model rotation curve found
and the circular velocities
$\delta=\sqrt{\frac{1}{N}\biggl(\sum_{i=1}^N
(V_{circ}(R_i)-\widetilde{V}_{circ}(R_i))^2\biggr)}$. As can be
seen, model III provides the best fit to the data; model II yields
a comparable result. For comparison, the last row gives the
residuals between our data and the model rotation curves from
Irrgang et al. (2013). It can be seen that the parameters we found
provide a more accurate fit, especially in the case of model III
(we managed to reduce the residual by a factor of 3).

Consider the first model separately. The mathematical description
of the halo potential in this model contains one more parameter,
the dimensionless coefficient $\gamma$. It was fixed at 2.02 in
the model by Allen and Santill\'an (1991); $\gamma=2.0$ in the
model by Irrgang et al. (2013). In the first version of fitting
the parameters of the potential, we also took $\gamma=2.0$. The
results obtained are presented in the first column of Table 1. We
see that the residual between the data and the constructed model
rotation curve is tangibly larger than that in models II and III.
In the second version of fitting, the coefficient $\gamma$ was
also included as a parameter being fitted. As a result, it turned
out that the minimum value of the residual for the first model for
our data was reached at $\gamma=6.3.$ The results of fitting other
parameters are presented in the second column of Table 1. We see
that the residual decreased noticeably.

To estimate the degree of uniformity of the residual noise (the
difference between the data and the model rotation curve), we used
the well-known concept of entropy calculated as follows:
$$
E=-\frac{1}{N}\sum_{i=1}^N |\Delta_i|\ln(|\Delta_i|),
$$
where $\Delta_i=V_{circ}(R_i)-\widetilde{V}_{circ}(R_i)$. The
higher the entropy, the more uniform the noise and, consequently,
the better the parameter fitting. Obviously, the combination of
$\delta$ and $E$ gives a more comprehensive idea of the quality of
fitting by various models than does $\delta$ alone. The entropy of
the residual noise is given in Table 1. As we see, model III
provides the greatest entropy of the noise, i.e., its uniformity.
Fitting the parameter $\gamma$ allowed the entropy of the residual
noise to be increased considerably in the first model as well.

Table 2 gives the physical quantities calculated from the derived
parameters of the model potentials (Eqs. (1)--(31)). These include
the local disk density $(\rho_\odot)_d$ (the local bulge density
is not given, because it is lower than the local disk density by
several orders of magnitude), the local dark matter density
$(\rho_\odot)_h,$ the local density of the entire matter
$\rho_\odot,$ the local surface density $\Sigma_{1.1}$ and
$\sum_{out}$, the two escape velocities from the Galaxy $V_{esc}$
(7) for $R=R_\odot$ and $R=200$~kpc, the linear circular rotation
velocity of the Sun $V_\odot,$ the Oort constants $A$ and $B$ from
Eqs. (8) and (9), and the Galactic mass $M_G$ for four radii of
the enclosing sphere. The Galactic rotation curves constructed for
all of the models under consideration are presented in Figs. 2 and
3.

Let us perform a comparative analysis of the constructed model
rotation curves.

In {\it model I,} the function describing the halo contribution to
the velocity curve is a nondecreasing one, as can be clearly seen
from Figs. 2 and 3. For this reason, the resulting model rotation
curve describes poorly the data already at distances $R$ greater
than 120 kpc, the Galactic mass estimate at $R\leq200$~kpc is
greatest compared to the remaining models in this paper (the last
rows in Table 2). However, it should be noted that model I at
$\gamma=6.3$ describes the data more accurately not only in the
inner Galaxy, especially on masers and especially at $R=2-8$ kpc
(only the rotation curve of model III (Fig. 3) provides a similar
fit to these data in this region), but also at great
Galactocentric distances than that at $\gamma=2.0$.

In {\it model II,} much of the mass in the inner Galaxy is
accounted for by dark matter. As can be seen from Table 1, the
lowest-mass central component corresponds to this model
($M_b=142\pm12~M_{g}$). At present, the amount of dark matter in
the inner Galactic region is being debated.
%

Whereas the presence of dark matter within the solar circle has
been solidly established by Iocco et al (2015), and many analysis
have aimed at the determination of its density profile in this
region with different methods (e.g. Catena and Ullio (2010), Iocco
et al. (2011), Bovy (2013), Pato et al. (2015), Pato and Iocco
(2015); see Read (2014) for a recent review), the innermost 2.5
kpc of the Galaxy is a more complicated region. In fact, the
presence of a bar and the departure from sphericity of the shape
of the bulge heavily hinder the reconstruction of dark matter
component, which is also expected to play a subleading role in the
total gravitational potential.

It can be concluded that although model II describes
satisfactorily the Galactic rotation curve in the $R$ range 0--200
kpc, it suggests the presence of a substantial dark matter mass in
the inner region of the Galaxy ($R<R_\odot$), which is most likely
far from reality.

{\it Model III} is currently one of the most commonly used models
(see, e.g., Sofue 2009; Kafle et al. 2012; Deason et al. 2012a).
In the outer Galaxy ($R>R_\odot$) its properties are similar to
those of model II, while in the inner Galaxy ($R>R_\odot$) the
dark matter mass is insignificant, which favorably distinguishes
this model from model II. As can be seen from the next to- last
row in Table 1, this model fits the data with the smallest
residual $\delta$ and the greatest entropy of the residual noise.

Such local parameters of the rotation curve as the velocity
$V_\odot$ and the Oort constants $A$ and $B$ are satisfactorily
reproduced by all three models considered.

Note that the escape velocities $V_{esc}~(R=200$~kpc) are usually
approximately half those at $R=R_\odot.$

As has already been said, as the target parameters in fitting
$\rho_\odot$ and $K_{z=1.1}/2\pi G$ we took the values justified
in Irrgang et al. (2013). Fitting these parameters basically led
to fitting the function $\rho_\odot(z)$ and, consequently, in
accordance with Eq. (33), the surface density $\Sigma_{1.1}$. The
results of fitting to the function $\rho_\odot(z)$ obtained in
Irrgang et al. (2013) and considered in this paper as a template
are shown in Fig. 4. As can be seen from the figures, all of the
models reproduce the required function $\rho_\odot(z)$ with a high
accuracy. Model I with $\gamma=6.3$ and model III provide the
highest accuracy.

\begin{figure}[p]
{\begin{center}
   \includegraphics[width=0.99\textwidth]{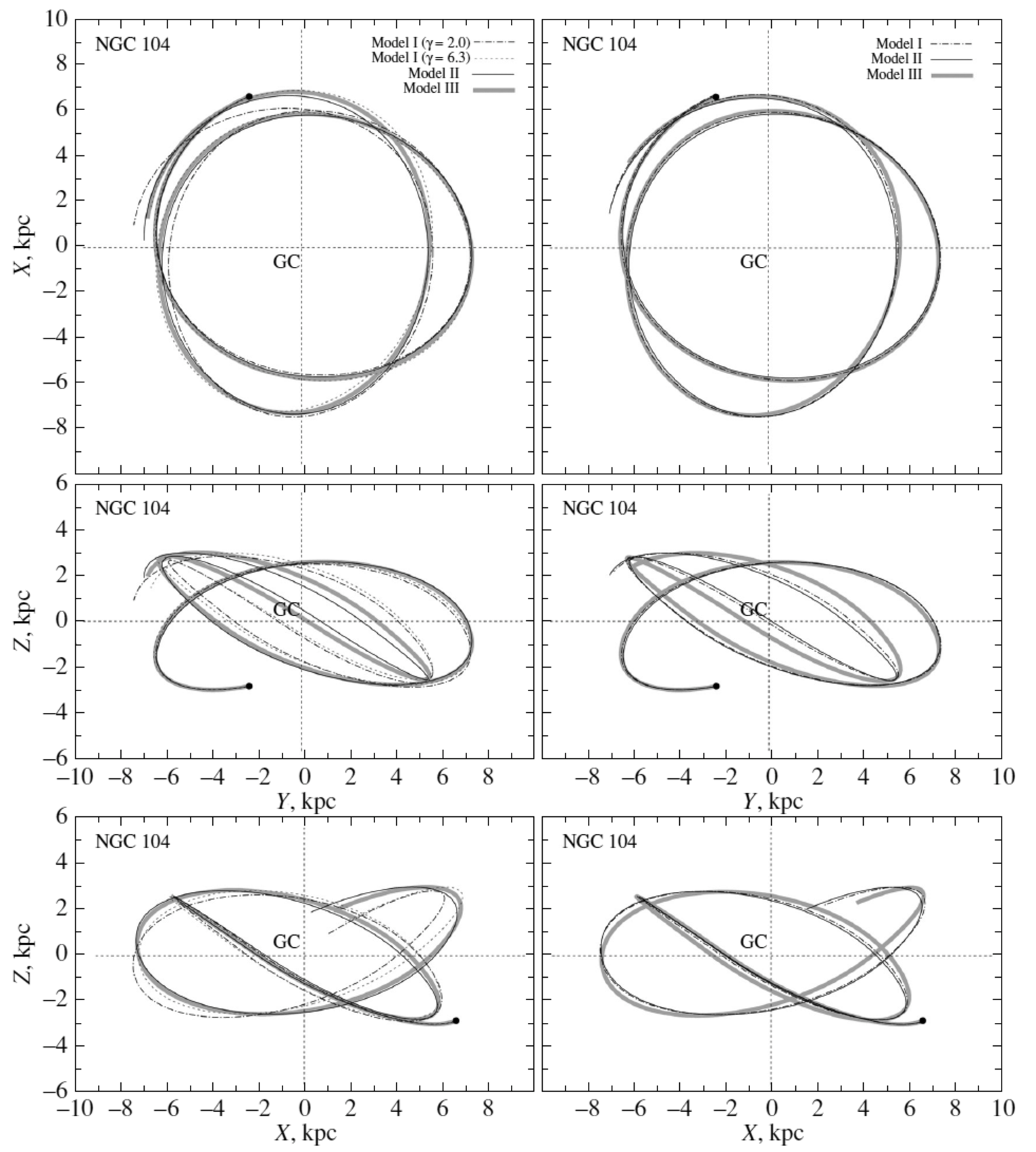}
 \caption{
Orbits of the globular cluster NGC 104 in a time interval of 400
Myr in the past for models I, II, and III with the parameters
derived in this paper (left) and the parameters from Irrgang et
al. (2013) (right). The filled circle corresponds to the beginning
of integration ($t=0$).
  } \label{f5}
\end{center}}
\end{figure}
\begin{figure}[p]
{\begin{center}
   \includegraphics[width=0.99\textwidth]{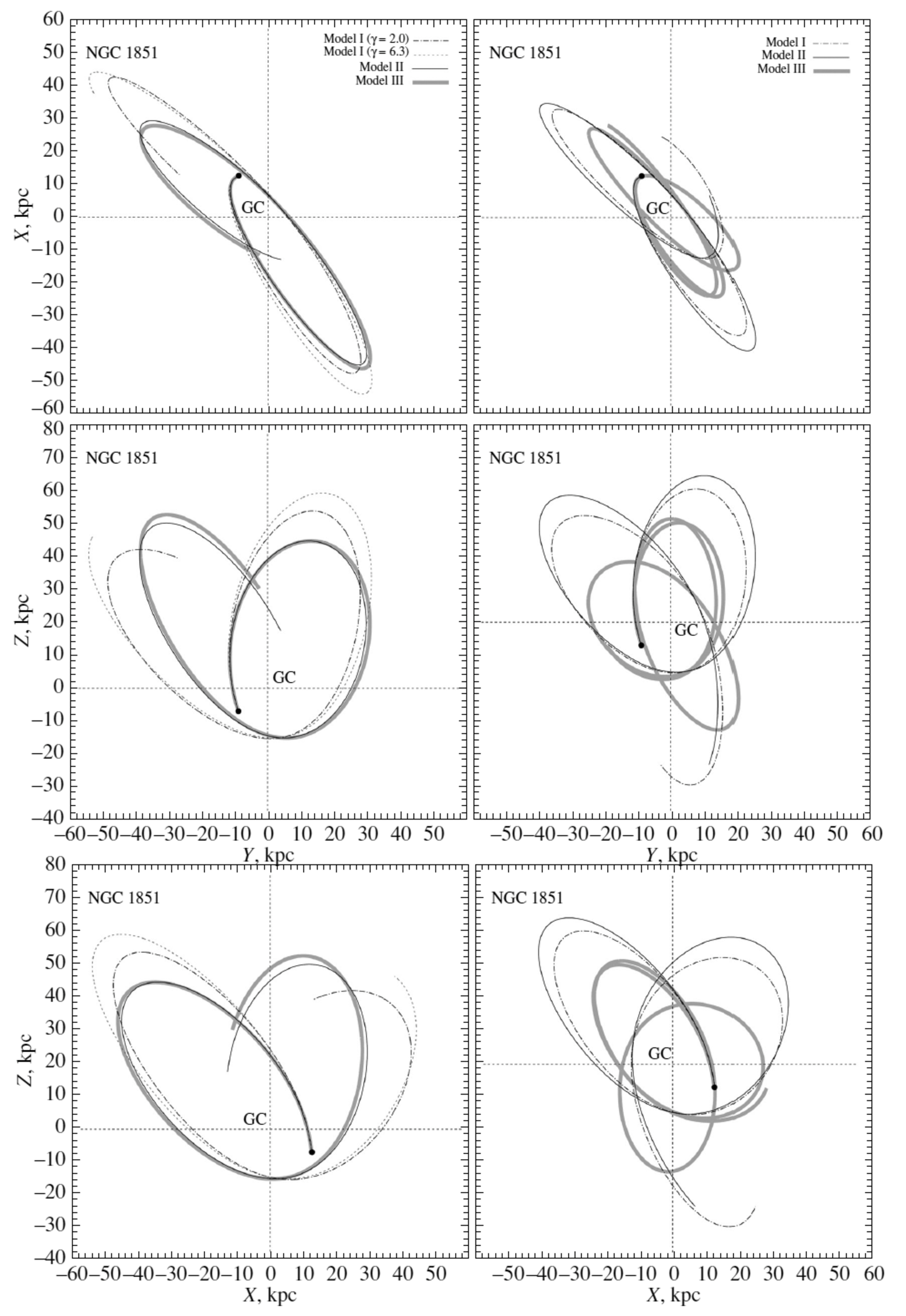}
 \caption{
Orbits of the globular clusterNGC1851 in a time interval of 2 Gyr
in the past for models I, II, and III with the parameters derived
in this paper (left) and the parameters from Irrgang et al. (2013)
(right). The filled circle corresponds to the beginning of
integration ($t=0$).
  } \label{f6}
\end{center}}
\end{figure}

\begin{figure}[t]
{\begin{center}
   \includegraphics[width=0.7\textwidth]{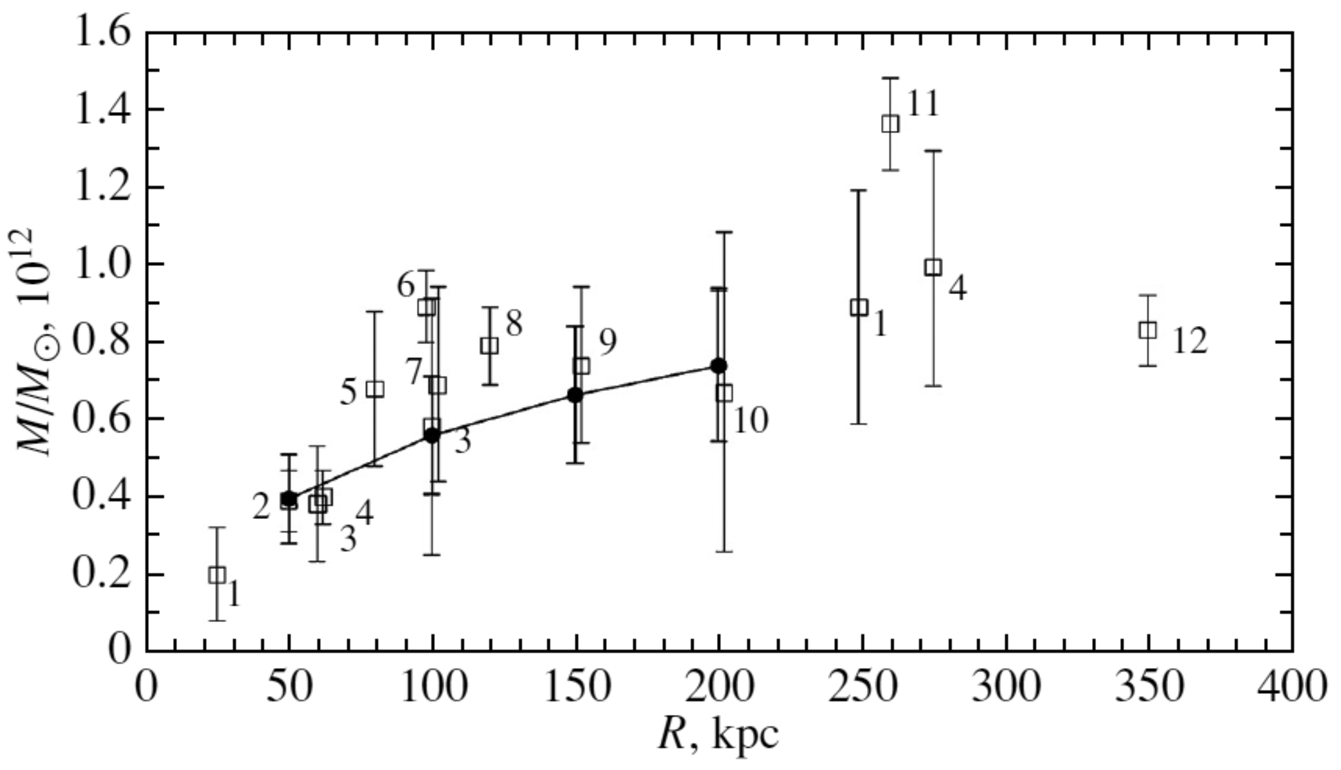}
 \caption{
The galactic mass estimates obtained by various authors (open
squares) and the estimates found in this paper based on model III
(thick line); the numbers indicate the following sources: 1--Kafle
et al. (2012), 2--Deason et al. (2012a), 3--Bhattacharjee et al.
(2013), 4--Xue et al. (2008), 5--Gnedin et al. (2010), 6--McMillan
(2011), 7--Dehnen and Binney (1998), 8--Battaglia et al. (2005),
9--Deason et al. (2012b), 10--Bhattacharjee et al. (2014),
11—Eadie et al. (2015), 12--Karachentsev et al. (2009).
  } \label{f7}
\end{center}}
\end{figure}

For an even subtler qualitative comparison of the constructed
model Galactic potentials between themselves and with the models
from Irrgang et al. (2013), we integrated the orbits of two
globular clusters into the past, NGC 104 and NGC 1851, in time
intervals of 400 Myr and 2 Gyr, respectively. The initial
distances, proper motions, and line-of-sight velocities of the
globular cluster NGC 104 were taken from Lane et al. (2012), where
the HST (Hubble Space Telescope) proper motion measurements are
provided. It should be noted that Table 1 from this paper gives
$\mu_\alpha \cos(\delta)$, not just $\mu_\alpha$. The data for the
globular cluster NGC 1851 were taken from the catalog by
Kharchenko et al. (2013). The peculiar velocity of the Sun
relative to the local standard of rest was taken to be
$(U,V,W)_\odot=(11.1, 12.24, 7.25)$ km s$^{-1}$ (Sch\"onrich et
al. 2010). We assumed the velocity $U$ to be directed toward the
Galactic center, $V$ to be in the direction of Galactic rotation,
and $W$ to be directed perpendicularly to the Galactic plane
toward the North Galactic Pole. The orbits of NGC 104 and NGC 1851
in three planes ($XY, XZ,$ and $YZ$) are presented in Figs. 5 and
6, respectively. To construct the orbits, we used the integrator
of the equations of motion given in the Appendix to the paper by
Irrgang et al. (2013) based on the fourth-order Runge–Kutta
algorithm.

The first object (NGC 104) is characterized by comparatively low
initial space velocities. In the Galactic coordinate system for
model III, $(U,V,W)=(-77.4, 187.2, 35.1)$ km s$^{-1}$. Its orbit
occupies a small region within $<8$~kpc of the Galactic center.
The second object (NGC 1851) has a considerably higher initial
velocity in Z coordinate. In the Galactic coordinate system,
$(U,V,W)=(-36.7, 102.5,-324.0)$ km s$^{-1}$. Its orbit encloses a
region with a radius up to 60 kpc over 2 Gyr. It was interesting
for us to trace how the orbits behaved in different models at
comparatively small and large Galactocentric distances.

The figures on the left and right panels correspond to our models
and the models from Irrgang et al. (2013), respectively. It can be
seen that the orbits of the first object corresponding to
different models are fairly close to one another. It should be
noted that our models I ($\gamma=6.3$), II, and III give almost
coincident orbits. The orbit corresponding to model I
($\gamma=2.0$) slightly differs, so that model I with $\gamma=6.3$
provides a closer orbit to the orbits from models II and III than
does model I with $\gamma=2.0$. As regards the models from Irrgang
et al. (2013), models I and II provide the closest orbits between
themselves.

The orbits from our models II and III for the second object, just
as for the first object, almost coincide. The orbits constructed
using models I differ noticeably. For the models with the
parameters from Irrgang et al. (2013), it can be seen that the
orbit from model III differs radically from the orbits of the two
other models. The orbits from models I and II also differ between
themselves, but only slightly. At the same time, they show a
similarity to the orbits from our models II and III.

Thus, by fitting the data at great Galactocentric distances, we
managed to significantly refine the parameters of all three models
considered here, especially model III from Navarro et al. (1997).
This enables a more accurate integration of the orbits of objects
with high space velocities on long time scales.

 \section*{DISCUSSION}
Let us compare the parameters of models I–III from Tables 1 and 2
with the estimates by Irrgang et al. (2013). In addition, it is
possible to compare the model rotation curves from model II using
Figs. 1 and 2. There are minimal differences between models II.
Our estimates of the mass parameters are always smaller. For
example, the halo mass for this model $M_h=69725~M_{g}$ from
Irrgang et al. (2013) is approximately a factor of 3 greater than
our value of $M_h=24572~M_{g}$. For model III the difference in
this parameter reaches already one order of magnitude.
Nevertheless, the estimates of the Galactic mass $M_G$ are quite
close at $R\leq50$~kpc and differ by a factor of 2 for model III.
At $R\leq200$~kpc the difference in $M_G$ estimates is a factor of
about 1.3 for model I, a factor of 2 for model II, and reaches a
factor of 4 for model III. Indeed, for model III we found
 $M_G=(0.75\pm0.19)\times10^{12}M_\odot,$ while Irrgang et al. (2013) estimated
 $M_G=(3.0\pm1.1)\times10^{12}M_\odot.$

Note once again that in Irrgang et al. (2013) the construction of
model potentials was based on the observations of masers with
measured trigonometric parallaxes located no farther than 20 kpc
from the Galactic center. Therefore, it is of interest to compare
our Galactic mass estimates with the results of other authors
obtained from objects far from the Galactic center.

Such a problem was solved by Sofue (2012) on the basis of
observational data in a wide range of distances $R:0-1000$~kpc
(Sofue 2009; Sofue et al. 2009, 2015). They are quite similar to
those data that were used by Bhattacharjee et al. (2014), because
they include the line-of-sight velocities of globular clusters and
dwarf galaxies. Based on a three-component model potential with
the dark matter halo represented in the NFW form (as in our model
III), Sofue (2012) estimated
 $M_G (R\leq385$~kpc$)=(0.70\pm0.10)\times10^{12}M_\odot.$
A different estimate of this author is
 $M_G (R\leq200$~kpc$)=(0.70\pm0.51)\times10^{12}M_\odot$ (Sofue 2015).
Note that Bhattacharjee et al. (2014) estimated
 $M_G (R\leq200$~kpc$)=(0.68\pm0.41)\times10^{12}M_\odot.$
Thus, there is good agreement with the result of our model III
 $M_G (R\leq200$~kpc$)=(0.75\pm0.19)\times10^{12}M_\odot.$

Xue et al. (2008) analyzed the line-of-sight velocities of blue
horizontal-branch giants at distances $R<60$~kpc. They constructed
a three-component model potential in which the dark halo mass was
represented in the NFW form, while the bulge and disk potentials
differed from those we used. These authors estimated the virial
mass of the Galaxy to be
 $M_G (R\leq R_{vir})=(1.0^{+0.3}_{-0.2})\times10^{12}M_\odot,$
where the virial radius was $R_{vir}=275^{+23}_{-20}$~kpc. Here we
can also see good agreement, within the error limits, with the
estimate obtained in this paper based on model III.

Based on the velocities of globular clusters and dwarf galaxies at
distances $R$ up to 260 kpc, Eadie et al. (2015) estimated the
Galactic mass to be $M_G
(R\leq260$~kpc$)=(1.37\pm0.07)\times10^{12}M_\odot.$

The Galactic mass within a sphere of radius 50 kpc,
 $M_G(R\leq50$~kpc$)\approx(0.41\pm0.12)\times10^{12}M_\odot,$
that we found based on three different model potentials is in good
agreement with the results of other authors. For example, Deason
et al. (2012a) estimate
 $M_G (R\leq50$~kpc$)=(0.42\pm0.04)\times10^{12}M_\odot,$; Williams and Evans (2015) found
 $M_G (R\leq50$~kpc$)=(0.45\pm0.15)\times10^{12}M_\odot$ from the velocities of horizontal
branch giants from the SDSS (Sloan Digital Sky Survey) catalogue.

Karachentsev et al. (2009) estimated the total mass of the Local
Group to be $(1.9\pm0.2)\times10^{12} M_\odot$ and the ratio of
the Galactic and M31 masses to be 4:5. The effect of local Hubble
flow deceleration and the distances and line-of-sight velocities
of galaxies in the neighborhoods of the Local Group were used for
this estimate. The total mass of the Galaxy obtained by this
independent method is
 $M_G(R\leq350$~kpc$)=(0.84\pm0.09)\times10^{12} M_\odot$, in complete
agreement with our estimates.

Figure 7 presents the Galactic mass estimates obtained by
different authors using different objects. Note that the results
marked by numbers 1 and 4 at $R>250$~kpc are the virial mass
estimates, while the direct estimates were obtained from the data
at $R<80$~kpc. On the whole, we can see good agreement of the
estimates found in this paper based on model III with the results
of different various.

 \section*{CONCLUSIONS}
We considered three three-component (bulge, disk, halo) model
Galactic potentials differing by the shape of the dark matter
halo. Present-day observational data spanning the range of
Galactocentric distances $R$ from 0 to $\sim$200 kpc were used to
refine the parameters of these models. We relied on the
line-of-sight velocities of hydrogen clouds at the tangent points
and the data on 130 masers with measured trigonometric parallaxes
up to distances of about 20 kpc and used the averaged rotation
velocities from Bhattacharjee et al. (2014) for greater distances.

In all of the models considered, the central component (bulge) and
the Galactic disk are represented in the form of Miyamoto and
Nagai (1975). The halo component is represented in the form of
Allen and Martos (1986) and Allen and Santill\'an (1991) in
model~I, in the form of Wilkinson and Evans (1999) in model~II,
and in the form of Navarro et al. (1997) in model~III.

For the Allen--Santill\'an model, a dimensionless coefficient
$\gamma$ has been included as a sought-for parameter for the first
time. In the traditional version of model I, $\gamma= 2.0.$ We
obtained $\gamma=6.3,$ which allowed the fit to the circular
velocity measurements by the model rotation curve to be improved
noticeably. Both versions of the potential were analyzed.

We fitted the model rotation curve to the rotation velocities of
Galactic objects known from observations by taking into account
the additional constraints on (a) the local matter density
$\rho_\odot$ and the force $K_{z=1.1}$ acting perpendicularly to
the Galactic plane. As a result, we obtained the model potentials
that described a stellar system consistent with the physical
characteristics of visible matter in the Galaxy known from
observations.

The Galactic mass within a sphere of radius 50 kpc,
 $M_G~(R\leq50$~kpv$)\approx(0.41\pm0.12)\times10^{12}M_\odot$,
was shown to satisfy all three models. The differences between the
models become increasingly significant with increasing radius $R.$
In model I, the Galactic mass within a sphere of radius 200 kpc at
$\gamma=2.0$ turns out to be greatest among the models considered,
 $M_G~(R\leq200$~kpc$)=(1.45\pm0.30)\times10^{12}M_\odot$,
 $M_G (R\leq200$~kpc$)=(1.29\pm0.14)\times10^{12}M_\odot,$ at $\gamma=6.3,$ and
the smallest value was found in model II,
 $M_G~(R\leq200$~kpc$)=(0.61\pm0.12)\times10^{12}M_\odot.$
In our view, model III (Navarro et al. 1997) is the best one among
those considered, because it ensures the smallest residual between
the data and the constructed model rotation curve provided that
the constraints on the local parameters hold with a high accuracy.
Here, the Galactic mass is
 $M_G~(R\leq200$~kpc$)=(0.75\pm0.19)\times10^{12}M_\odot.$

A comparative analysis with the models by Irrgang et al. (2013),
including those using the integration of orbits for the two
globular clusters NGC~104 and NGC~1851 as an example, has been
performed. The third model is shown to have been subjected to a
significant improvement.

 \subsection*{ACKNOWLEDGMENTS}
We are grateful to the referees for their helpful remarks that
contributed to an improvement of this paper. This work was
supported by the ``Transitional and Explosive Processes in
Astrophysics'' Program P--41 of the Presidium of Russian Academy
of Sciences.

 \bigskip{REFERENCES}\medskip
 {\small

 1. C. Allen and M.A. Martos, Rev. Mex. Astron. Astrofis. 13, 137 (1986).

 2. C. Allen and A. Santill\'an, Rev. Mex. Astron. Astrofis. 22, 255 (1991).

 3.~G. Battaglia, A. Helmi, H. Morrison, P. Harding, E.W. Olszewski, M. Mateo,
    K.C. Freeman, J. Norris, and S.A. Shectman, Mon. Not. R. Astron. Soc. 364, 433 (2005).

 4. P. Bhattacharjee, S. Chaudhury, S. Kundu, and S. Majumdar, Phys. Rev. D 87, 083525 (2013).

 5. P. Bhattacharjee, S. Chaudhury, and S. Kundu, Astrophys. J. 785, 63 (2014).

 6. O. Bienaym\'e, B. Famaey, A. Siebert, K.C. Freeman, B.K.
Gibson, G. Gilmore, E.K. Grebel, J. Bland- Hawthorn, et al.,
Astron. Astrophys. 571, AA92 (2014).

 7. V.V. Bobylev and A.T. Bajkova, Astron. Lett. 39, 809 (2013).

 8. V.V. Bobylev and A.T. Bajkova, Astron. Lett. 40, 389 (2014).

 9. J. Bovy, {\it Probes of Dark Matter on Galaxy Scales},
 AAS Topical Conference Series Vol.~1. Proc. of conf. 14--19 July 2013 in Monterey, CA.
 Bull. American Astron. Soc., 45, \#7, \#402.01 (2013).

 10. J. Bovy and H.-W. Rix, Astrophys. J. 779, 115 (2013).

 11. R.A. Burns, H. Imai, T.Handa, T. Omodaka, A. Nakagawa, T.
 Nagayama, and Y. Ueno, Mon. Not. R. Astron. Soc. 453, 3163 (2015).

 12. W.B. Burton and M.A. Gordon, Astron. Astrophys. 63, 7 (1978).

 13.~R. Catena, and P. Ullio, J. Cosmol. Astropart. Phys. 8, 4 (2010).

 14. Y.K. Choi, K. Hachisuka, M.J. Reid, Y. Xu, A. Brunthaler, K.
     M. Menten, and T.M. Dame, Astrophys. J. 790, 99 (2014).

 15. A.J. Deason, V. Belokurov, N.W. Evans, and J. An,
     Mon. Not. R. Astron. Soc. 424, L44 (2012a).

 16. A.J. Deason, V. Belokurov, N.W. Evans, S.E. Koposov, R.J.
Cooke, J. Penarrubia, C.F.P. Laporte, M. Fellhauer, et al., Mon.
Not. R. Astron. Soc. 425, 2840 (2012b).

 17. W. Dehnen and J. Binney, Mon. Not. R. Astron. Soc. 294, 429 (1998).

 18. G.M. Eadie, W.E. Harris, and L.M. Widrow, Astrophys. J. 806, 54 (2015).

19. H. Edelmann, R. Napiwotzki, U. Heber, N. Christlieb, and D.
Reimers, Astrophys. J. 634, L181 (2005).

20. S. Garbari, C. Liu, J. I. Read, and G. Lake, Mon. Not. R.
Astron. Soc. 425, 1445 (2012).

 21. E. Gardner, P. Nurmi, C. Flynn, and S. Mikkola, Mon. Not. R. Astron. Soc. 411, 947 (2011).

 22. O.Y. Gnedin, W.R. Brown, M.J. Geller, and S.J. Kenyon, Astrophys. J. 720, L108 (2010).

 23. A.O. Gromov, I.I. Nikiforov, and L.P. Ossipkov, arXiv:1508.05298 (2015).

 24. K. Hachisuka, Y.K. Choi, M.J. Reid, A. Brunthaler, K.M.
Menten, A. Sanna, and T.M. Dame, Astrophys. J. 800, 2 (2015).

 25. J. Holmberg and C. Flynn, Mon. Not. R. Astron. Soc. 313, 209 (2000).

 26. J. Holmberg and C. Flynn, Mon. Not. R. Astron. Soc. 352, 440 (2004).

 27. L.G. Hou, J.L. Han, and W.B. Shi, Astron. Astrophys. 499, 473 (2009).

 28.~F. Iocco, M. Pato, G. Bertone, and P. Jetzer,
 J. Cosmol. Astropart. Phys. 11, 29 (2011).

 29. F. Iocco, M. Pato, and G. Bertone, Nat. Phys. 11, 245 (2015).

 30. A. Irrgang, B. Wilcox, E. Tucker, and L. Schiefelbein, Astron.
Astrophys. 549, 137 (2013).

31. R.R. Kafle, S. Sharma, G.F. Lewis, and J. Bland-Hawthorn,
Astrophys. J. 761, 98 (2012).

32. I.D. Karachentsev, O.G. Kashibadze, D.I. Makarov, and R.B.
Tully, Mon. Not. R. Astron. Soc. 393, 1265 (2009).

33. N.V. Kharchenko, A.E. Piskunov, E. Schilbach, S. R\"oser, and
R.-D. Scholz, Astron. Astrophys. 558, A53 (2013).

34. V.I. Korchagin, T.M. Girard, T.V. Borkova, D.T. Dinescu, and
W.F. van Altena, Astron. J. 126, 2896 (2003).

35. G.G. Kuzmin, Astron. Zh. 33, 27 (1956).

36. R.R. Lane, A.H.W. K\"{u}pper, and D.C. Heggie, Mon. Not. R.
Astron. Soc. 423, 2845 (2012).

 37. S. L\'epine, A. Koch, R.M. Rich, and K. Kuijken, Astrophys. J. 741, 100 (2011).


 38. C.F. McKee, A. Parravano, and D.J. Hollenbach, Astrophys. J. 814, 13 (2015).

 39. P.J. McMillan, Mon. Not. R. Astron. Soc. 414, 2446 (2011).

 40. M. Miyamoto and R. Nagai, Publ. Astron. Soc. Jpn. 27, 533 (1975).

 41. K. Motogi, K. Sorai, M. Honma, T. Hirota, K. Hachisuka, K. Niinuma, K. Sugiyama,
     Y. Yonekura, and K. Fujisawa, arXiv:1502.00376 (2015).

 42. J.F. Navarro, C.S. Frenk, and S.D.M. White, Astrophys. J. 490, 493 (1997).

 43. M. Pato and F. Iocco, Astrophys. J. 803, 3 (2015).

 44. M. Pato, F. Iocco, and G. Bertone, J. Cosmol. Astropart. Phys. 12, 1 (2015).

45. E.-M. Pauli, R. Napiwotzki, U. Heber, M. Altmann, and M.
Odenkirchen, Astron. Astrophys. 447, 173 (2006).

46. C.B. Pereira, E. Jilinski, N.A. Drake, D.B. de Castro, V.G.
Ortega, C. Chavero, and F. Roig, Astron. Astrophys. 543, A58
(2012).

47. H.C. Plummer, Mon. Not. R. Astron. Soc. 71, 460 (1911).

48. J.I. Read, J. Phys. G: Nucl. Part. Phys. 41, f3101 (2014).

49. M.J. Reid, K.M. Menten, A. Brunthaler, X.W. Zheng, T.M.Dame,
Y. Xu, Y.Wu, B. Zhang, et al., Astrophys. J. 783, 130 (2014).

50. A. Sanna, M.J. Reid, K.M. Menten, T.M. Dame, B. Zhang, M.
Sato, A. Brunthaler, L.Moscadelli, and K. Immer, Astrophys. J.
781, 108 (2014).

51. M. Sato, Y.W. Wu, K. Immer, B. Zhang, A. Sanna, M.J. Reid, T.
M. Dame, A. Brunthaler, and K.M. Menten, Astrophys. J. 793, 72
(2014).

52. R. Sch\"onrich, J. Binney, and W. Dehnen, Mon. Not. R. Astron.
Soc. 403, 1829 (2010).

 53. Y. Sofue, Publ. Astron. Soc. Jpn. 61, 153 (2009).

 54. Y. Sofue, M. Honma, and T. Omodaka, Publ. Astron. Soc. Jpn. 61, 227 (2009).

 55. Y. Sofue, Publ. Astron. Soc. Jpn. 64, 75 (2012).

 56. Y. Sofue, Publ. Astron. Soc. Jpn. 67, 75 (2015).

57. J.S. Urquhart, M.G. Hoare, S.L. Lumsden, R.D. Oudmaijer, T.J.
T. Moore, J.C. Mottram, H.D.B. Cooper, M. Mottram, and H.C.
Rogers, Mon. Not. R. Astron. Soc. 420, 1656 (2012).

58. M.I. Wilkinson and N.W. Evans, Mon. Not. R. Astron. Soc. 310,
645 (1999).

59. A.A. Williams and N.W. Evans, Mon. Not. R. Astron. Soc. 454,
698 (2015).

60. Z.-Y. Wu, J. Ma, X. Zhou, and C.-H. Du, Astron. J. 141, 104
(2011).

61. Y.W. Wu, M. Sato, M.J. Reid, L. Moscadelli, B. Zhang, Y. Xu,
A. Brunthaler, K.M. Menten, T.M. Dame, and X.W. Zheng, Astron.
Astrophys. 566, 17 (2014).

62. Y. Xu, J.J. Li, M.J. Reid, K.M. Menten, X.W. Zheng, A.
Brunthaler, L. Moscadelli, T.M. Dame, and B. Zhang, Astrophys. J.
769, 15 (2013).

63. X.X. Xue, H.-W. Rix, G. Zhao, P. Re Fiorentin, T. Naab, M.
Steinmetz, F.C. van den Bosch, et al., Astrophys. J. 684, 1143
(2008).

64. L. Zhang, H.-W. Rix, G. van de Ven, J. Bovy, C. Liu, and G.
Zhao, Astrophys. J. 772, 108 (2013).

65. The Hipparcos and Tycho Catalogues, ESA SP- 1200 (1997).

 }
\end{document}